# Big Data Is not just a New Type, but a New Paradigm

Bin Jiang

Faculty of Engineering and Sustainable Development, Division of GIScience
University of Gävle, SE-801 76 Gävle, Sweden
Email: bin.jiang@hig.se

There is no doubt that we have entered an unprecedented big data era. But what is big data? While this question has been answered in many ways (e.g., Mayer-Schonberger and Cukier 2013), a commonly cited definition involves characterizing big data with several 'Vs', such as volume, variety, veloctiy, and veracity (Laney 2001). Big data has had enormous impacts on science, particularly social science, in the 21$^{st}$ century (Lazer et al. 2009, Watts 2007). Geographic information consititutes a very important type of big data, thanks to geospatial technologies such as the global positioning system, geographic information systems, and remote sensing. Geospatial technologies have dramatically enabled social media, the Internet, and mobile devices, which have led to large amounts of georeferenced locations about things and people being made available for better understanding environment and urban systems.

Large amounts of volunteered geographic information (VGI), created by volunteers through crowdsourcing, represent a new phenomenon that has arisen from Web 2.0 technologies (Goodchild 2007, Sui et al. 2013). VGI constitutes one of the most important types of user-generated content and is quickly becoming a new type of asserted geographic information that complements the traditional authoritative geographic information collected by governmental agencies or private organizations. In the context of this special issue, VGI is broadly defined as referring to any georeferenced information that is freely distributable without copyright constraints to anyone who is interested. Georeferenced big data (hereafter, big data) such as Flickr photos and their locations, digital tickets and travel cards, tweets and their locations, are not just a new type of data, but a new paradigm. In what follows, we discuss some of our personal reflections on big data; for example: how it differs from small data, some of the fundamental ways of thinking behind big data, and how we should develop new insights into big data.

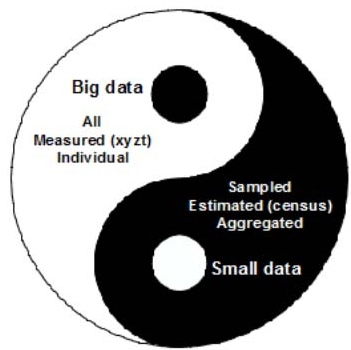

Figure 1: Differences between big and small data

Big data differs fundamentally from small data in terms of the data's characteristics (Mayer-Schonberger and Cukier 2013). Small data are mainly sampled (for example, census or statistical data), while big data are automatically harvested using, for example, crawling techniques or application programming interface provided by social media providers, from a large population of users or crowds. Given the large population, big data are often referred to as all rather than a small part of it. Put another way, small data is like an elephant seen by a blind man, while big data is the elephant itself. Small data, such as census data, are essentially estimated, while big data such as VGI and social media data are measured accurately at very fine spatial and temperal resolutions. Big data are defined at an individual level rather than an aggregated level (as small data are). These three distinguishing characteristics – all, measured, and individual – make big data big and differentiate them from their small data counterparts (Figure 1). Using the long tail theory (Anderson 2006), big data can be characterized as the tail – a majority – while small data is the head – a minority. This head-and-tail analogy can be widely seen in examples such as national telecoms vs. Skype, Encyclopedia Britannica vs. Wikipedia, national mapping agencies vs. OpenStreetMap, and governments/CNN vs. WikiLeaks. The differences between big and small data imply two different ways of thinking: (1) that the former is bottom-up and decentralied, and (2) that the latter is top-down and centralized.

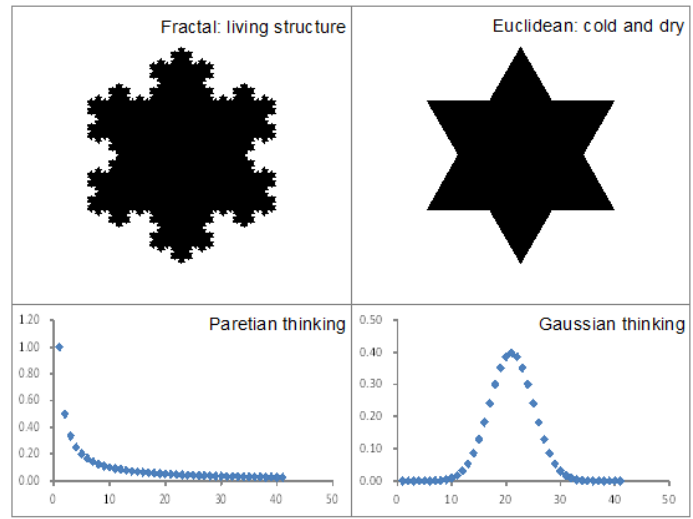

Figure 2: Different paradigms for big and small data analytics
(Note: 'Cold and dry', and 'living structure' are credited to Mandelbrot (1982), and Alexander (2002–2005), respectively)

Big data differs fundamentally from small data in terms of data analytics, both geometrially and statistically (Figure 2). Euclidean geometry has been a default paradigm for geographic representation and serves as the scientific foundation for various geographic concepts such as distances, directions, and map projections. It is not surprising that many geospatial analyses are Euclidean, as they essentially deal with regular shapes and smooth curves. However, geographic features are essentially irregular and rough, which means that Euclidean geometry is not an appropriate paradigm for geography, even in the small data era. Instead, fractal geometry (Mandelbrot 1982) should be adopted for developing new insights into geograpic forms and dynamics (e.g., Batty and Longley 1994), particularly in the context of big data (Jiang 2015a). Statistically, geographic features are very diverse and heterogeneous, so they are likely to demonstrate a heavy-tailed distribution that lacks a characteristic mean; these are referred to as scale-free. In this connection, a Paretian way of thinking or power law statistics is more appropriate for dealing with the heterogeneity (Jiang 2015b). In the small data era, a Gaussian way of thinking (or Gaussian statistics) was widely adopted based on the assumption that geographic features can be characterized by a well-defined mean. Below, we use the concept of natural cities to elaborate on the different paradigms and their implications for big -data analytics.

Big data are crowdsourced by very diverse and heterogeneous individuals – so-called crowds. The locations of these crowds are diverse, independent, and decentralized, which means that aggregation of the massive amount of locations is likely to lead to some wisdom of crowds (Surowiecki 2004). For example, natural cities emerge from the bottom up, or from a massive number of social media users' locations (Jiang and Miao 2015). Based on a huge triangulated irregular network (TIN) consisting of the locations of all tweets in a country or the world, those locations with short edges (shorter than an average of all the edges of the TIN) are aggregated to create natural cities. The natural cities provide a counterpart for the cities that are conventionally imposed by authorities from the top down. Cities are mainly created for legal and adminstrative purposes, and there is hardly a universal standard for all cities in the world. A city in Sweden may not qualify as a city in China due to its size. The lack of a universal standard for defining cities is the reason why many previous efforts to delineate city boundaries based on remote sensing imagery have failed. In other words, these efforts failed not because of techniques or methods, but because of the definitions of cities or the way of thinking. Conventional definitions of cities are a product of the small-data era and are very useful for administration and management. On the other hand, natural cities are produced from big data and are unique and of critical importance for studying city structure and dynamics (Jiang and Miao 2015, Jiang 2015a).

Night-time images accurately record lights emitted from human settlements at night on the Earth's surface, and can also be used to derive natural cities. A night-time image consists of millions of pixels, each of which is valued between 0 and 63 gray-scales. These images have been previously calibrated by removing unrelated information such as wildfires and lights coming from the sea, leaving only stable lights coming from human settlements or cities. The massive number of pixels can be accounted as crowds, the average value of which can be used as a meaningful cutoff for deriving natural cities for the globe. This average value plays the same role as the average edge in the TIN in the earlier example. The natural cities derived from the night-time image are found to exhibit Zipf's law strikingly at the global scale. In this connection, big data is not so much about bigness but completeness, since all the natural cities on the Earth act as a whole. Not only city sizes, but city numbers are found to follow Zipf's law. The city numbers in individual countries are inversely proportional to their country ranks; the number of cities in the largest country is twice as many as in the second-largest city, and three times as many as in the third-largest country, and so on (Jiang et al. 2015). One of the major advantages of working with big data is that it enables us to see something that is hard to see with small data.

Due to its diversity and heterogeneity, big data inevitably demonstrates the scaling pattern of far more small things than large ones; for example, far more short edges than long ones in the TIN, or far more dark/gray pixels than light ones in the night-time images. The large and small things constitute the head and the tail of a heavy-tailed distribution, respectively, in the rank-size plot (Zipf 1949). Importantly, the scaling pattern of far more small things than large ones recurs multiple times, again and again for the things in the head. This recurring scaling pattern is what underlies the new definition of fractals based on head/tail breaks as both a classification scheme (Jiang 2013) and a visualization tool (Jiang 2015a) for data with a heavy-tailed distribution. More specifically, the head/tail breaks divide a whole around an average into a few large things in the head and many small things in the tail, and continue recursively for the division for the head until the notion of far more small things than large ones is violated. The head is self-similar to the whole, the very notion of self-similarity of all kinds of fractals. We believe that the head/tail breaks, or Paretian and fractal thinking in general have profound implications for big data and big data analytics.

**Acknowledgment**
This paper is a first draft of the introduction to the special issue on volunteered geographic information published in Computers, Environment and Urban Systems (2015, 53, 1–122). I would like to thank my colleague Dr. Jean-Claude Thill, who expanded the draft towards a broader scope.